\documentstyle[11pt,aaspp4,flushrt]{article}

\newcommand\psr{PSR~J0437$-$4715}
\newcommand\ntts{PSR~B1937+21}
\newcommand\etal{{\it et al.}}
\def\lapp{\ifmmode\stackrel{<}{_{\sim}}\else$\stackrel{<}{_{\sim}}$\fi}
\def\gapp{\ifmmode\stackrel{>}{_{\sim}}\else$\stackrel{>}{_{\sim}}$\fi}

\begin{document}

\title{Radio Pulse Properties of the Millisecond Pulsar
\psr. I. Observations at 20 cm}

\author{F. A. Jenet}
\affil{California Institute of Technology, Space Radiation Laboratory,
Pasadena, CA 91125}
\authoremail{merlyn@fezzik.caltech.edu}

\author{S. B. Anderson}
\affil{California Institute of Technology, Space Radiation Laboratory,
Pasadena, CA 91125}
\authoremail{sba@jelly.caltech.edu}

\author{V. M. Kaspi\altaffilmark{1}}
\affil{Department of Physics and Center for Space Research, Massachusetts
Institute of Technology, 37-621, Cambridge, MA 02139}
\authoremail{vicky@space.mit.edu}

\author{T. A. Prince}
\affil{California Institute of Technology, Space Radiation Laboratory,
Pasadena, CA 91125}
\authoremail{prince@caltech.edu}

\author{S. C. Unwin\altaffilmark{2}}
\affil{California Institute of Technology, Space Radiation Laboratory,
Pasadena, CA 91125}
\authoremail{scu@jelly.caltech.edu}

\altaffiltext{1}{Hubble Fellow}
\altaffiltext{2}{Current Address: Jet Propulsion Laboratory, Mail Stop 306-388, 4800 Oak Grove Drive, Pasadena, California 91109-8099}

\begin{abstract}

We present a total of 48~minutes of observations of the nearby, bright
millisecond pulsar \psr~taken at the Parkes radio observatory in
Australia.  The data were obtained at a central radio frequency of
1380~MHz using a high-speed tape recorder that permitted coherent
Nyquist sampling of 50 MHz of bandwidth in each of two polarizations.
Using the high time resolution available from this voltage recording
technique, we have studied a variety of single-pulse properties, most
for the first time in a millisecond pulsar.  We show that individual
pulses are broadband, have pulse widths ranging from $\sim$10~$\mu$s
($\sim 0.6^{\circ}$ in pulse longitude) to $\sim$300~$\mu$s ($\sim
20^{\circ}$) with a mean pulse width of $\sim$~65$\mu$s ($\sim
4^{\circ}$), exhibit a wide variety of morphologies, and can be highly
linearly polarized.  Single pulse peaks can be as high as 205~Jy (over
$\sim$40 times the average pulse peak), and have a probability
distribution similar to those of slow-rotating pulsars.  We observed
no single pulse energy exceeding $\sim$4.4 times the average pulse
energy, ruling out ``giant pulses'' as have been seen for the Crab and
\ntts\ pulsars.  \psr\ does not exhibit classical microstructure or
show any signs of a preferred time scale that could be associated with
primary emitters; single pulse modulation has been observed to be
consistent with amplitude-modulated noise down to time scales of 80
ns. We observe a significant inverse correlation between pulse peak
and width. Thus, the average pulse profile produced by selecting for
large pulse peaks is narrower than the standard average profile.  We
find no evidence for ``diffractive'' quantization effects in the
individual pulse arrival times or amplitudes as have been reported for
this pulsar at lower radio frequency using coarser time resolution
(\cite{amdv97}).  Overall, we find that the single pulse properties of
\psr\ are similar to those of the common slow-rotating pulsars, even
though this pulsar's magnetosphere and surface magnetic field are
several orders of magnitude smaller than those of the general
population.  The pulsar radio emission mechanism must therefore be
insensitive to these fundamental neutron star properties.

\end{abstract}


\section{Introduction} \label{sec:intro}

Single pulse studies of millisecond pulsars are considerably more
difficult to perform than are those of slow pulsars.  Faster data
rates are needed to study millisecond pulsars with comparable pulse
phase resolution, and finer radio frequency resolution is required to
minimize the effect of interstellar dispersion.  Also, interstellar
scattering time scales comparable to the pulse duration render
studying individual pulse morphologies impossible, while steep
millisecond pulsar spectra preclude observations of sufficient
sensitivity at higher radio frequencies where scattering is less
important.

Yet single pulse studies of millisecond pulsars are highly desirable
for two reasons.  First, the origin of the radio emission that has
made isolated neutron stars famous is, even 30 years after their
discovery, still a mystery. The high brightness temperatures ($\sim
10^{30}$K) associated with the radio emission point to coherent
processes which are poorly understood even under less exotic
conditions (\cite{mel96}). Previous observations of slow pulsars have
not constrained the emission mechanism sufficiently; the study of
radio emission properties of millisecond pulsars may provide important
new clues. Millisecond pulsars, because of their fast spin periods,
have much smaller light-cylinder radii, and hence magnetospheres, than
slow pulsars.  They also have lower surface magnetic field strengths
than the general pulsar population (most likely resulting from their
having been ``recycled'' by a binary companion through the accretion
of mass and angular momentum). Were the radio emission mechanism at
all dependent on such properties, millisecond pulsars should have
different radio properties than the slower-spinning general
population.  The second reason single pulse studies of millisecond
pulsars are important is that millisecond pulsar timing is well-known
to be an unparalleled source of precision astrometric and
astrophysical information.  Among factors possibly limiting timing
precision is the stability of the average profile, which depends on
the properties of single pulses.

Only recently has a systematic study of single pulses from millisecond
pulsars become possible, largely due to improving computational and data
recording technologies.  To date, the only such investigation has been
for the 1.5~ms pulsar \ntts\ (\cite{wcs84,sb95,bac95,cstt96}).
Interestingly, \ntts\ exhibits giant radio pulses like those seen
elsewhere only in the Crab pulsar.  With the single pulse
properties of only one millisecond pulsar having been studied in any
detail, the question of whether  all millisecond pulsars show
similar properties naturally arises.  Unfortunately, \ntts\ suffers
interstellar scattering at time scales comparable to the duration of a
single pulse at the radio frequencies at which it has been observed,
rendering detailed study of individual pulse morphologies difficult.

Here we report on high-time-resolution single pulse studies of a
second millisecond pulsar, \psr.  This pulsar's large flux density and
low dispersion measure (DM), and the corresponding scarcity of
line-of-sight scattering material, render it an obvious target for
single pulse work.  Some single pulse investigations of \psr\ have
been reported (\cite{jlh+93,amdv97}) but none have had sufficient time
resolution to resolve most individual pulses. Using a fast recording
device and powerful supercomputers, we have been able to resolve all
pulses in our data, the narrowest being $\sim$10~$\mu$s. 


\section{Observations and Analysis} \label{sec:obs}

\subsection{Parkes Observations} \label{ssec:parkes}

All data reported on here were obtained at the 64-m radio telescope of
Parkes Observatory in New South Wales, Australia on 24 and 25 July,
1995 at a central radio frequency of 1380~MHz.  Observations were made
using a cryogenically-cooled, dual-channel system which received
orthogonal linear polarizations.  The signal path from the receiver
was as follows.  A local oscillator signal, locked to the Parkes
Observatory frequency standard, was amplified and then mixed with the
incoming RF signal from the receiver and low-pass filtered to form a
single-sideband IF.  These IF signals (one for each polarization) were
relayed to the control room, where they were amplified and mixed with
a second LO.  This second mixer operated as a baseband quadrature
mixer, with the in-phase and quadrature-phase output signals low-pass
filtered to 25 MHz.  Thus, a total bandwidth of 50 MHz at center
frequency 1380 MHz was available for recording.  The four analog
signals from the complex downconverter were then digitized with 2-bit
resolution and recorded at the Nyquist rate (400 Mbit~s$^{-1}$) on a
digital tape recorder (Datatape LP-400) along with timing information
synchronized to the Parkes Observatory clock.  The output bus operates
at 50 Msample~s$^{-1}$, and is 8 bits wide: 2 digitizer bits $\times$
2 polarizations $\times$ 2 signal phases.  Thus, a single 13~min
observation resulted in 39~Gbytes of raw voltage data.  Detailed
information about the baseband mixing, digitizing, and data recording
system is reported by Jenet \etal\ (1997).  \nocite{jcpu97}

The epochs and durations of the observations of \psr\ reported on in
this paper are summarized in Table~\ref{ta:obs}.  A calibration scan
was done prior to Observations 1 and 3 by moving the telescope
off-source and pulsing a noise diode source at a fixed
frequency.  The amplitude of the calibration source, and hence the
absolute flux scale in Janskys for our pulsar observations were
determined using observations of the bright radio continuum source Hydra.

\subsection{Data Analysis} 
\label{ssec:analysis}

Storage of raw voltage data permits great flexibility in data
analysis, but also requires significant computational power. Most data
reduction was therefore done on the massively-parallel Caltech
512-node Intel Paragon XPS L38 supercomputer which has peak
computation rate of 38.4 Gflop~s$^{-1}$, as well as on the 32-node Intel
Paragon XPS A4 supercomputer which has 4.3 Gflop~s$^{-1}$ peak computation
rate.  Data from the Datatape recorder were read into a 0.5 Gbyte
Datatape variable rate buffer, and then onto the supercomputer file
systems via a high-speed HIPPI network.  Most analysis was done using
specialized parallel code written in C++, with the NX message passing
interface.  For details about the hardware and analysis software
tools, see Jenet \etal\ (1997).

\subsubsection{Dedispersion} \label{sssec:dedisp}

Once on the parallel computers, the raw data is unpacked into floating
point numbers and the interstellar dispersion handled by one or both
of two techniques.  The first, and the less computationally intensive,
uses a software ``incoherent filter-bank,'' which simulates the output
of the conventional hardware filter-bank spectrometer, namely samples
of power in many individual narrow frequency channels.  The
channelized data are subsequently added with appropriate time delays
to achieve a dedispersed time series having time resolution limited by
residual dispersion within channels.  The second, more
computationally-demanding dedispersion method does a direct Fourier
deconvolution of the interstellar medium transfer function, i.e.
``coherent dedispersion'' (\cite{hr75}). For \psr, it is easy to show
that 128 filterbank frequency channels are needed to minimize
single-channel dispersion and time resolution.  This results
in a time resolution of 2.56 $\mu s$ and a dispersion smearing of
approximately 3.26 $\mu s$. In principle, coherent dedispersion can
yield a time resolution equal to the inverse of twice the bandwidth,
however in practice one is limited by the precision with which the DM
is known.  For \psr, the uncertainty on the published DM
(\cite{sbm+97}) implies that the true time resolution of our
coherently dedispersed time series is no better than $\sim$200~ns; we
conservatively chose to average our coherently dedispersed time series
to a time resolution of 320~ns.  Higher time resolutions, when
necessary, were achieved by coherently dedispersing a smaller
bandwidth, obtained using a coarse-resolution software filter bank.
For example, we achieved 80~ns resolution by coherently dedispersing a
6.25~MHz band. It is important to note that multi-frequency DM
measurements are very accurate, but not necessarily precise. Therefore,
such DM values may be of limited use in single pulse studies.

The presence of large amplitude signals in the 2-bit quantized voltage
data will introduce unwanted artifacts in the final dedispersed time
series if proper care is not taken. For bright pulsars like \psr,
appropriate corrections are crucial for proper pulse morphology
analyses. Two major quantization effects have been identified: power
underestimation and power scattering. We have minimized the power
underestimation effects at the data unpacking stage by dynamically
adjusting the assigned voltage levels. The effects of the scattered
power have only been corrected in the incoherently dedispersed data. A
more detailed discussion of these quantization effects and the
algorithms used to correct them may be found elsewhere (Jenet 1997, in
preparation).

Thus, we produced a (coherently or incoherently) dedispersed time
series for each of the two polarizations.  The mean value calculated
over a megasample was then subtracted in each polarization, and a gain
factor was applied to convert the raw data values to Janskys.  The
channels were then added to yield the total intensity, or Stokes I.
Faraday rotation in the interstellar medium causes a time delay
between opposite circular polarizations, as well as a rotation of the
angle of linear polarization across the band, however given the low DM
and rotation measure (\cite{nms+97}), both effects are negligible
here.

\subsubsection{Folding the data with the pulse period} \label{ssec:fold}

\begin{figure}
\plotone{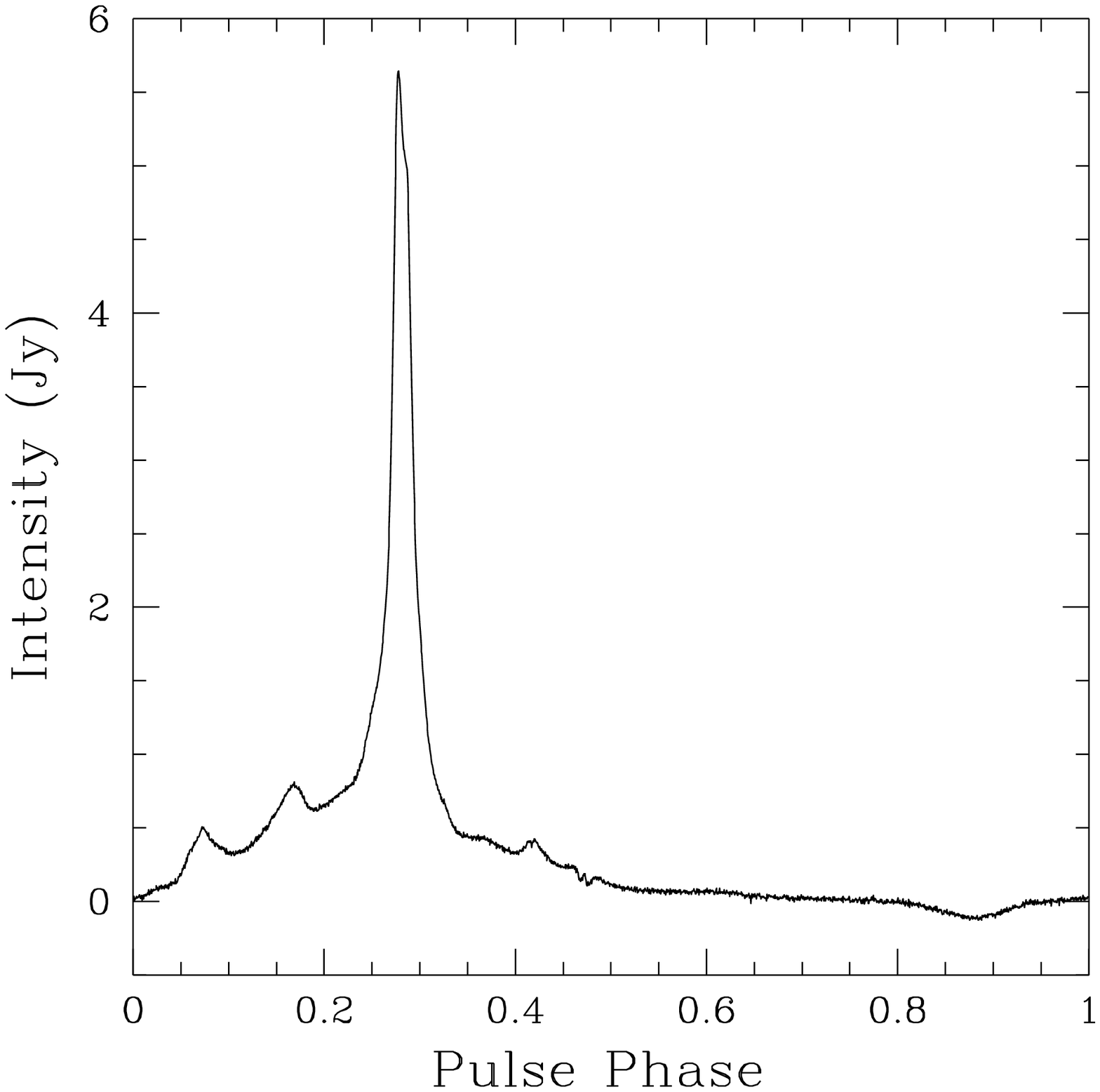}
\figcaption[avgpulse.eps]{Average pulse profile obtained by folding data from
Observation 3 (see Table~\protect\ref{ta:obs}).  There are 2048 bins across 
the pulse profile; dispersion smearing in the finite-width simulated
filters is 3.26~$\mu$s, just larger than one bin.  The small ``dip''
seen near phase 0.9 is an instrumental effect. \label{fig:avgpulse}}
\end{figure}

The known pulsar ephemeris (\cite{sbm+97}; provided for convenience in
Table~\ref{ta:parms}) was used with the TEMPO software package
(\cite{tw89}) to calculate the expected topocentric pulse period and
the pulse phase once every 671~ms. The average pulse profile, shown in
Figure~\ref{fig:avgpulse}, was obtained by dedispersing and folding
the data in Observation 3, assuming a dispersion measure (DM) of
2.64515~pc~cm$^{-3}$. In the Figure there are 2048 bins across the
period, and remaining dispersion smearing due to the finite size of
the simulated filters is 3.26~$\mu$s, just larger than one bin.  The
pulse morphology is identical to that seen by other groups
(e.g. \cite{mj95,nms+97}) except for the presence of a broad, shallow
``dip'' in the baseline. A spectral analysis shows that the dip is not
dispersed. Hence it is an instrumental artifact. Since each pulse of
emission last only about 400 $\mu$s in our 50 MHz bandpass, the dip,
which is about 2 ms away from the main peak, should not affect the
remainder of the analysis. We suspect that this dip is the result of
some non-linear process occurring in the receiver chain when the pulse
is outside of our bandpass although we have not yet seen this effect
in other pulsars observed with the same receiver chain.

\section{Results and Discussion} \label{sec:results}

\subsection{General Features of Single Pulses}

\begin{figure}
\plotone{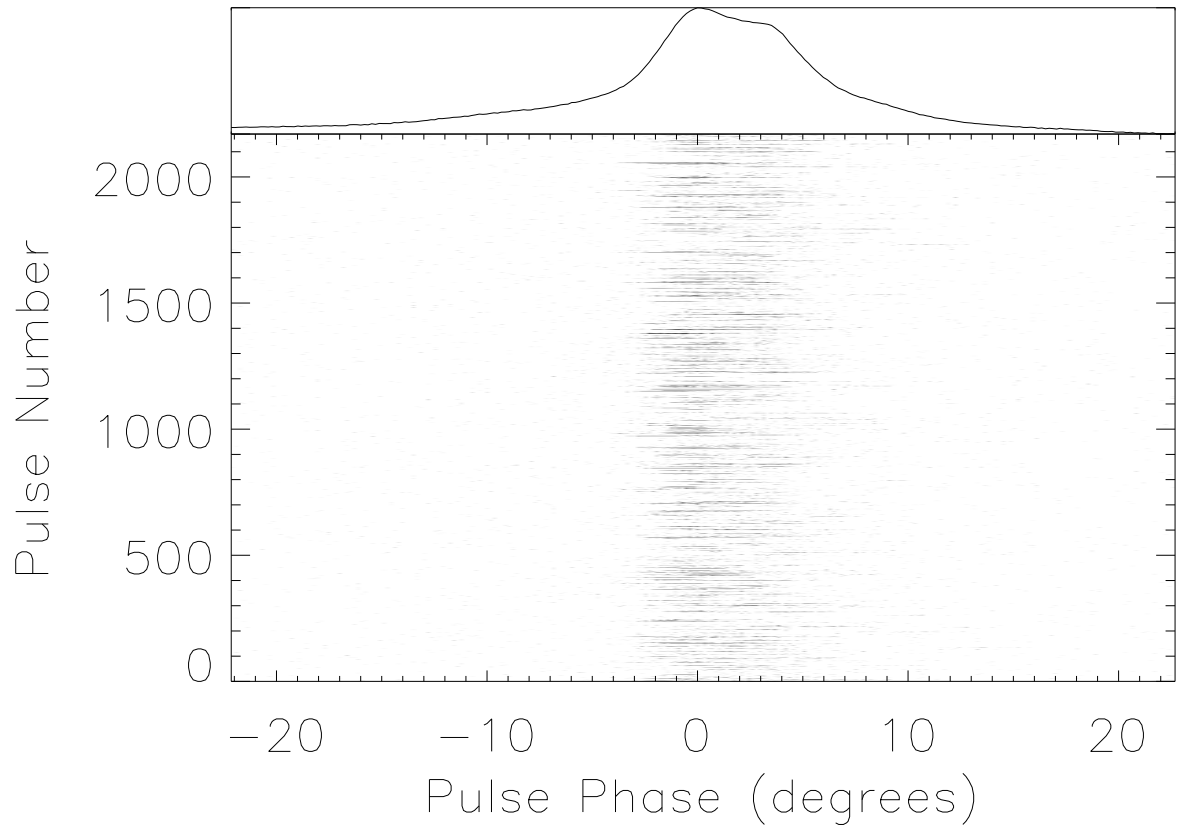}
\figcaption[greyscale.eps]{Grey scale plot showing single pulses of
\psr\ during a single $\sim$13-s observation.  Note that only the
central $\sim$0.7~ms are shown; the average profile formed by these
pulses is shown at the top. \label{fig:greyscale}}
\end{figure}

\begin{figure}
\plotone{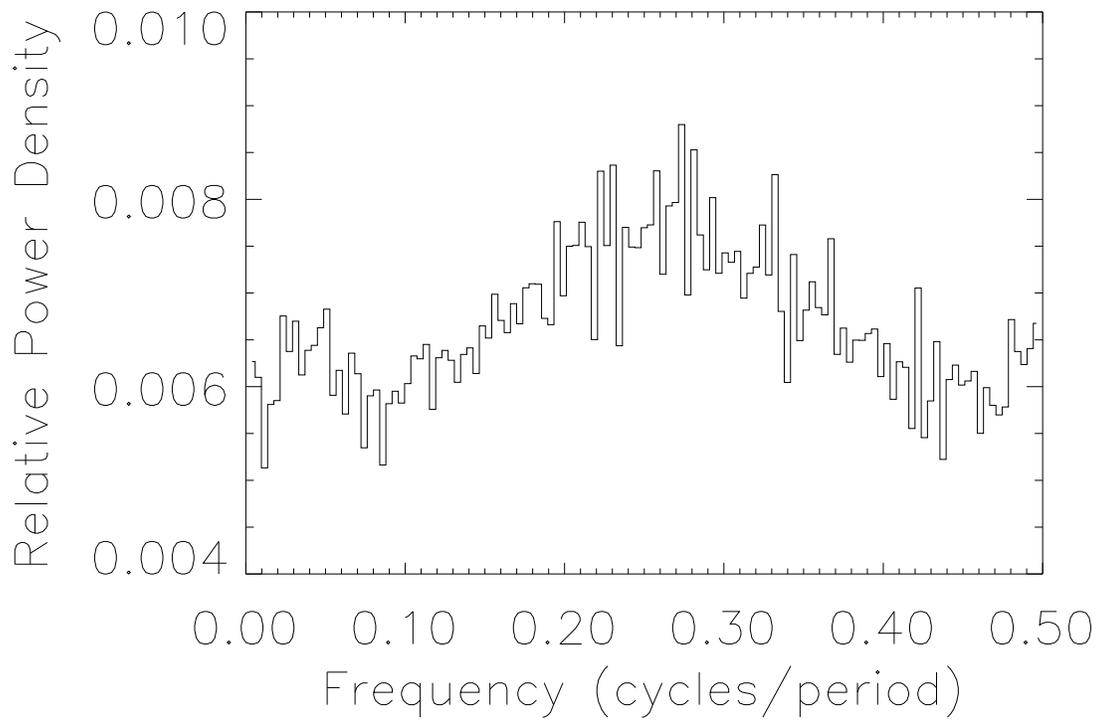}
\figcaption[fluc.eps]{The average intensity fluctuation spectrum
calculated for 100,352 pulses at zero pulse phase (see text). The
power density spectrum is normalized by the zero frequency
power. The frequency axis is normalized by the pulsar frequency.\label{fig:fluc}}
\end{figure}

\begin{figure}
\plotone{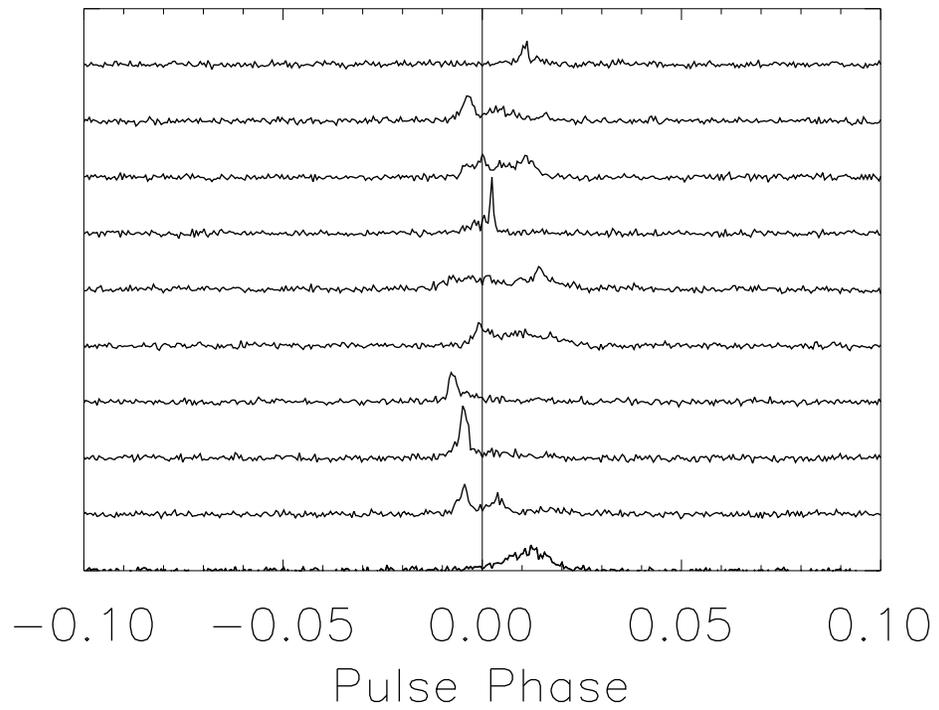}
\figcaption[roguesg.eps]{A random sampling of phase-aligned individual
bright pulses from \psr, using a software filter bank.  The time
resolution in this plot is 2.56~$\mu$s, with 3.26~$\mu$s of DM
smearing. \label{fig:roguesg}}
\end{figure}

Figure~\ref{fig:greyscale} is a greyscale plot of 12.5-s of data phase
aligned with the pulse peak.  Only 368~$\mu$s on either side of the
main peak is shown.  Data used for this plot are incoherently
dedispersed, corresponding to a time of resolution 2.56~$\mu$s with
3.26~$\mu$s DM smearing (as for Fig.~\ref{fig:avgpulse}). No evidence
for drifting subpulses can been seen in this
figure. Figure~\ref{fig:greyscale} does show regions where the pulse
is absent, but we found that averaging as little as ten of these
regions reveals the presence of a pulse. Hence, we have no evidence
for pulse nulling. The average fluctuation spectrum shown in
Figure~\ref{fig:fluc} displays a broad feature indicating a
correlation or ``memory'' between pulses. The fluctuation spectrum is
calculated by first extracting a time series of intensities at a fixed
pulse phase and then Fourier transforming this time series
(\cite{bac73}). In order to reduce the variance in the spectrum, 392
consecutive 256 point spectra were averaged together. The fluctuation
spectrum shown in Figure~\ref{fig:fluc} was calculated at zero pulse
phase. Structure in the fluctuation spectrum is common among slow
pulsars (\cite{mt77}, \cite{bac73}). Figure~\ref{fig:roguesg} shows a
``rogues gallery'' of several phase-aligned sub-pulses from the same
data set that produced Figure~\ref{fig:greyscale}. It is clear from
Figures \ref{fig:greyscale} and \ref{fig:roguesg} that, as is typical
of pulsars, the average profile is a sum of many individual subpulses,
which, although forming a relatively stable average profile,
individually exhibit a variety of morphologies.

\begin{figure}
\plotone{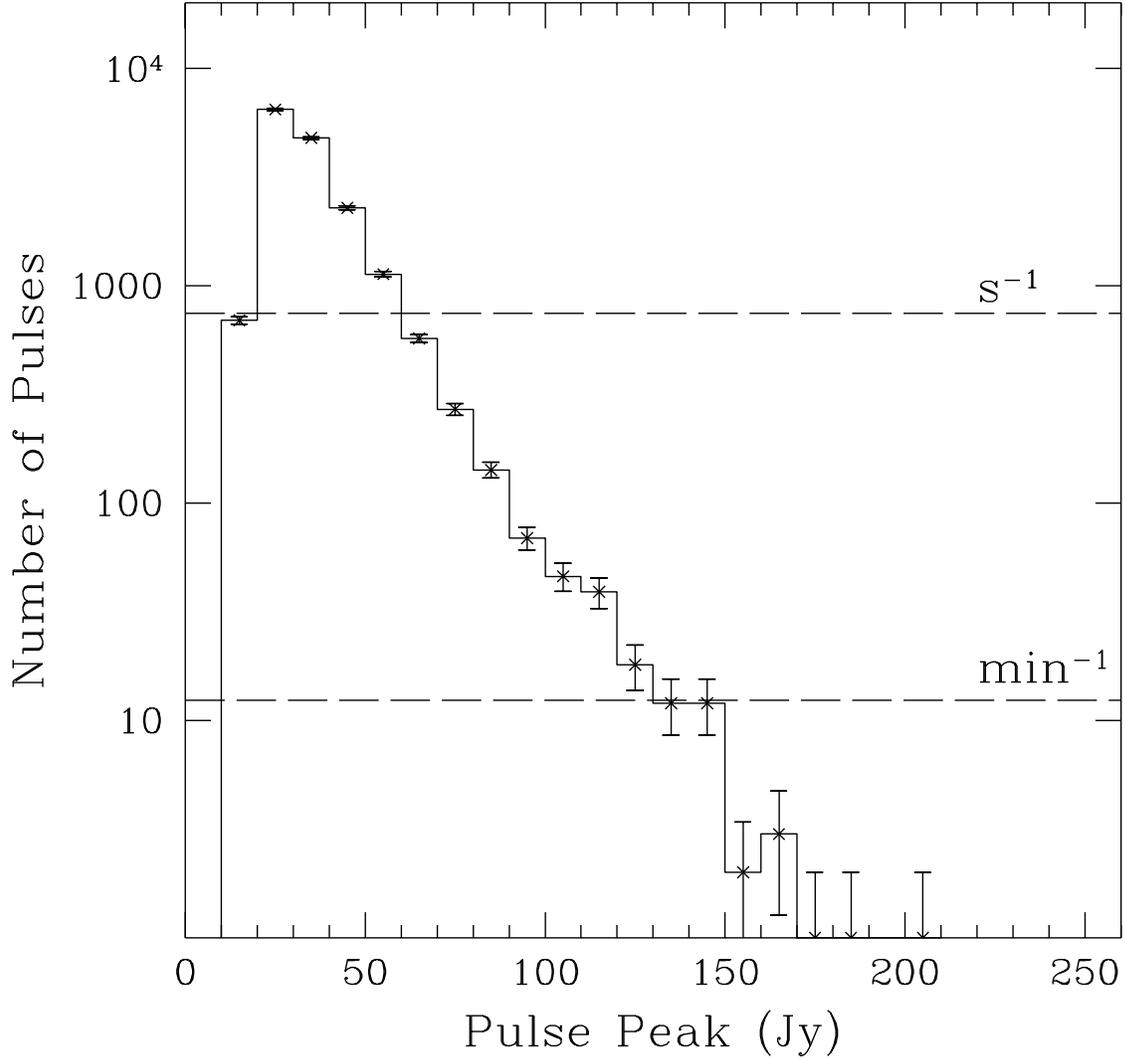}
\figcaption[plghist.eps]{Histogram of peak pulse amplitudes in Jy from
Observation 3 (see Table~\protect\ref{ta:obs}), dedispersed using a
software filter bank. The time resolution was 2.56 $\mu$s. Rates for
two peak values (1 per second and 1 per minute) are indicated by dashed lines. The low flux-density
cutoff is due to an imposed threshold criterion. \label{fig:peakhist}}
\end{figure}

Figure~\ref{fig:peakhist} shows a semi-logarithmic histogram of
sub-pulse peaks in units of Janskys for Observation 3.\footnote{The
mean fluxes in the different observations were different because of
scintillation.  For simplicity we choose to report results from
Observation 3, in which the pulsar was brightest.}  Results for our
other observations are similar.  In Observation 3, the mean pulse peak
was 5.6~Jy.  From Figure~\ref{fig:peakhist}, the largest peaks in
Observation~3 were $\sim$40 times the mean value. We have included only sub-pulses that are statistically
significant,i.e., the chances of randomly obtaining the measured on pulse
power must have been less then 1 in $10^{11}$ assuming gaussian
statistics.
The low flux-density cutoff therefore represents our sensitivity
threshold given the receiver noise temperature.  Only 20\% of pulses
satisfied this criterion.  The distribution for this 12.4~min
observation is well-described over most of the observed pulse peak
range by the expression,
\begin{equation}
\log_{10} ({\rm N}) = a + b \times {\rm P}_{peak},
\label{eq:peakhist}
\end{equation}
where $a = 4.51 \pm 0.01$, $b = -0.0262 \pm 0.0002$~Jy$^{-1}$, ${\rm
P}_{peak}$ is the peak power in Jy, and N is the number of pulses.
However extrapolating this expression to a pulse peak of 205~Jy
suggests only 0.14 pulses at this height should have been observed;
that one at 205~Jy was seen suggests the distribution may have a
high-peak tail that longer integrations might make observable.

\begin{figure}
\plotone{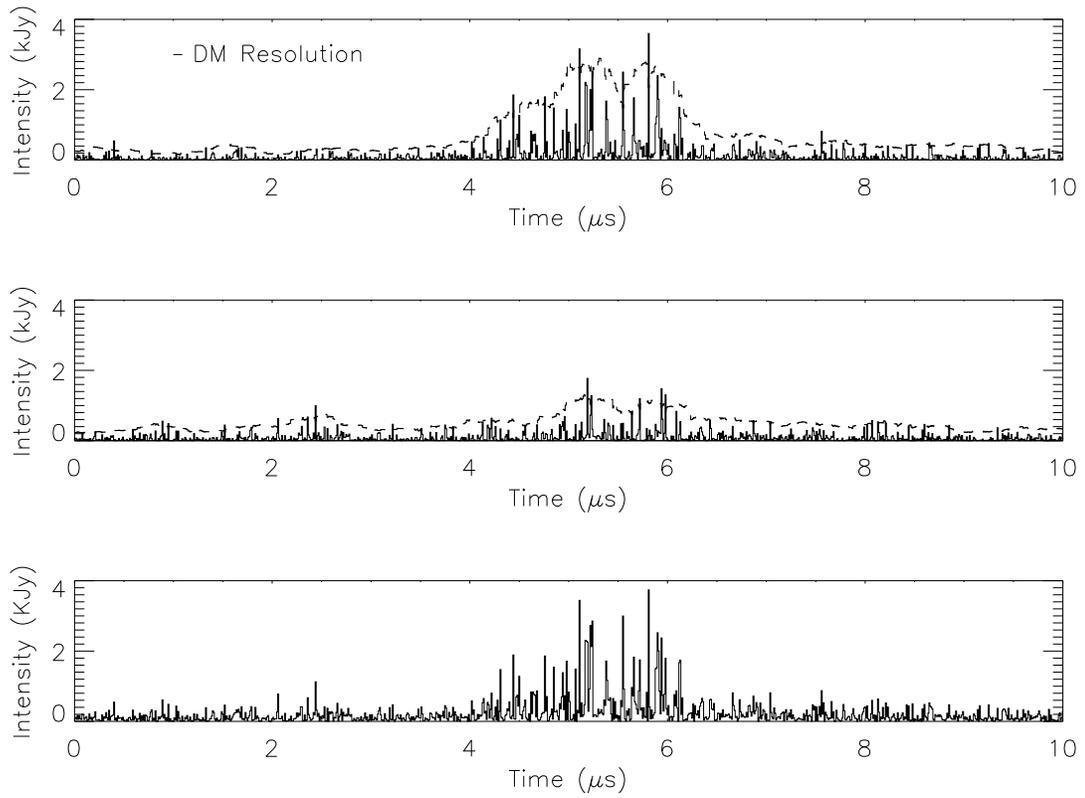}
\figcaption[largestpulse.eps]{The largest amplitude pulse in our data. The time resolution
is intrinsically 10ns but is increased to 200 ns by DM uncertainties
and non-ideal filter response. The dashed line represents the 98\%
confidence level (see text). \label{fig:largestpulse}}
\end{figure}

Figure~\ref{fig:largestpulse} shows a large amplitude subpulse
coherently dedispersed with an intrinsic time resolution of 10 ns. Two
factors limit our actual time resolution to $\sim$ 200~ns: DM
uncertainties and non-ideal low pass filter response. The low pass filters
used in the downconverter have a group delay of up to 80 ns at the
upper band edge. Neither of these effects would hide the presence of a
coherent signal: although they would make a noise signal appear more like a
gaussian noise signal, they would not turn a coherent signal into
a random noise signal. The envelopes superimposed on the two linear
polarizations represent the 98\% confidence levels for $\chi_1^2$
statistics. If the statistics of the voltage data were given by a
gaussian distribution, we would expect the power to be given by a
$\chi_1^2$ distribution. Using 25 data points to either side of the
point in question, we calculated the mean power and used this to find
the value of the power such that the probability of being less then
that value is 0.98. As can be seen from this figure, the statistics of
the emission are consistent with modulated noise. It is also clear
from Figure \ref{fig:largestpulse} that this single pulse is
substantially linearly polarized. The largest ten pulses in our data all
show similar behaviour.

\begin{figure}
\plotone{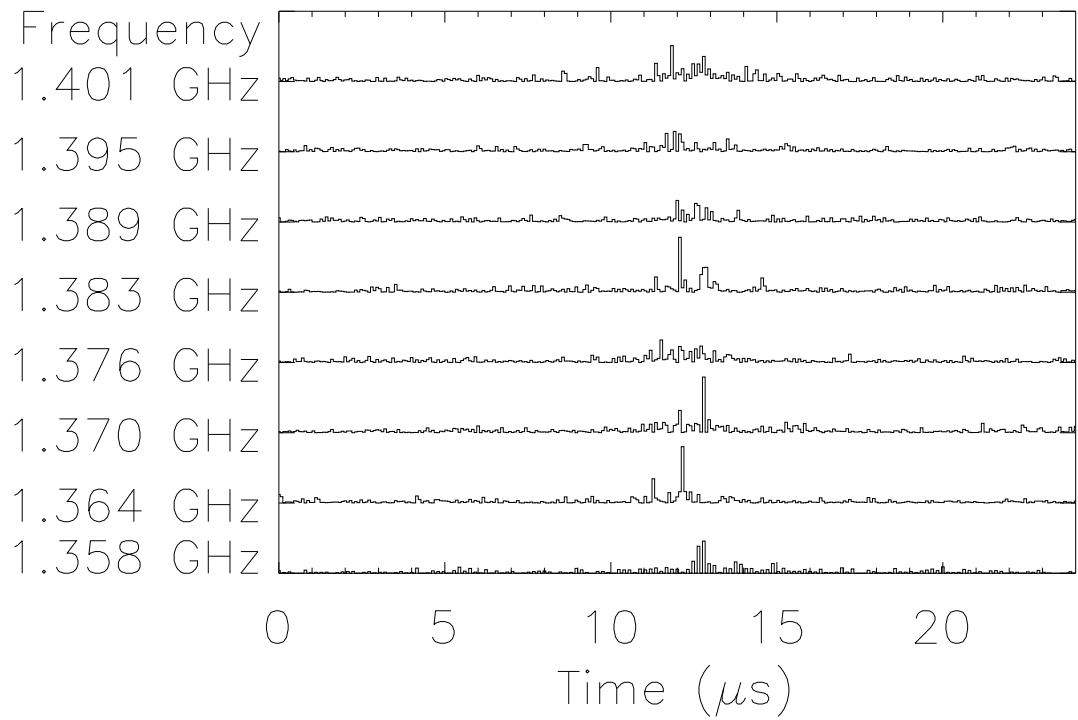}
\figcaption[waterfall.eps]{The largest amplitude pulse in our data, in
each of eight frequency sub-bands.  Here the data were coherently
dedispersed, then subjected to an eight channel software filterbank at
zero DM.  The lowest frequency channel , the nyquist channel, appears
to have a lower signal-to-noise ratio because its statistics are
intrinsically different from the other channels.\label{fig:waterfall}}
\end{figure}

The same subpulse is shown in Figure~\ref{fig:waterfall} in each of
eight radio frequency sub-bands.  It is clear that this pulse is
broad-band, extending at least over 50~MHz bandwidth.  The
scintillation bandwidth at this frequency is much larger than 50~MHz
so is not relevant.  The sub-band structure is consistent with a
gaussian noise process.

\begin{figure}
\plotone{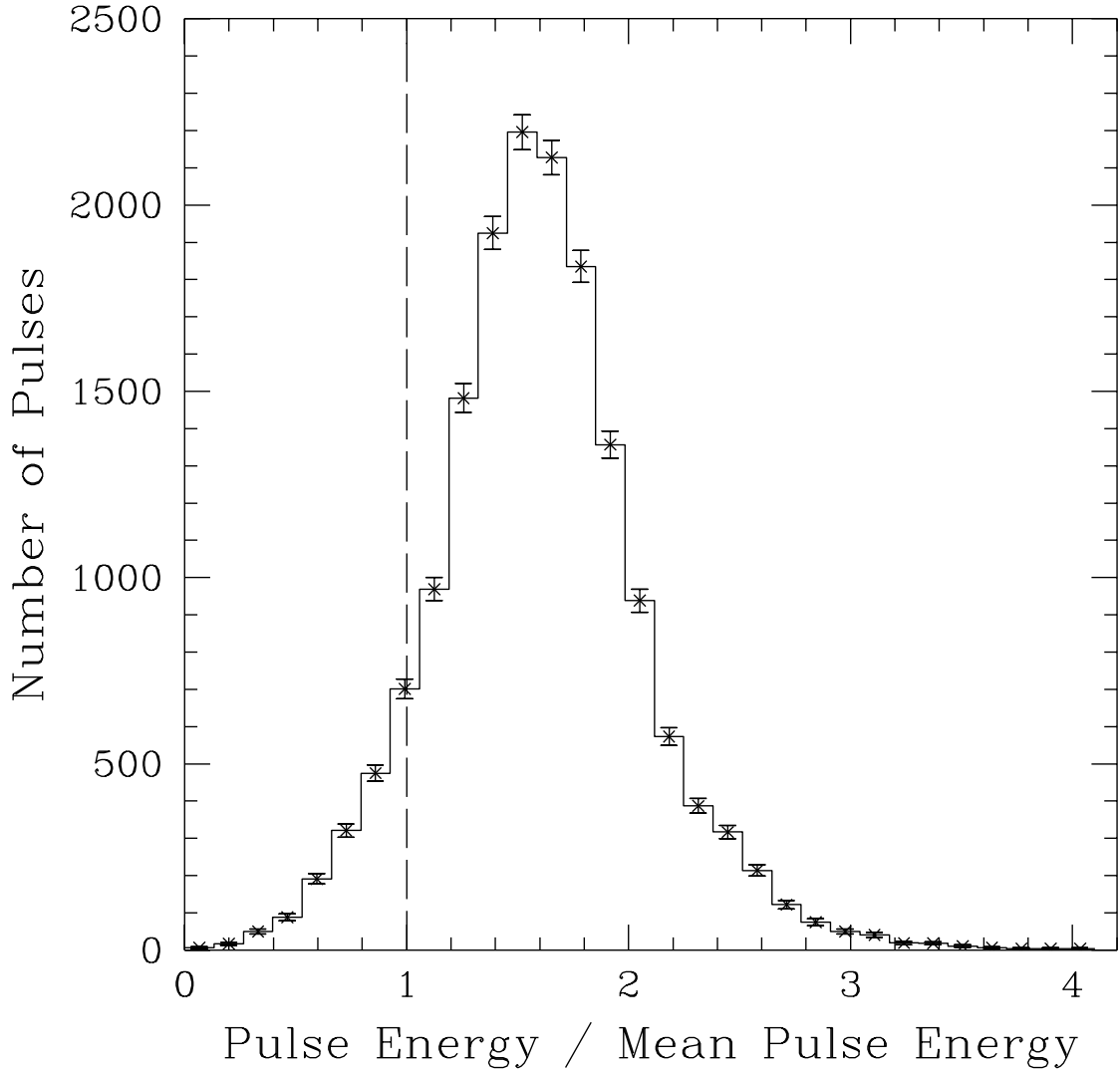}
\figcaption[elghist.eps]{Histogram of pulse energies for Observation
3, including statistically significant pulses only.  The vertical
dashed line is the mean pulse energy, which falls below the peak of
the distribution, since the mean pulse is below the noise level.  The
largest pulse energy we observed was 4.4 times the
mean.\label{fig:energyhist}}
\end{figure}

``Giant pulses'' from the millisecond pulsar \ntts\ have been observed
(\cite{wcs84,sb95,bac95,cstt96}), where ``giant'' implies single
pulses having energies much larger than the average single pulse
energy.  To investigate whether the large pulses seen in PSR
J0437$-$4715 are also ``giant'' in this sense, we produced a histogram
of pulse energies; this is shown in Figure~\ref{fig:energyhist} for
Observation 3.  Results for the other observations are similar. Only
statistically significant sub-pulses were included.  The vertical
dashed line indicates the mean energy.  Its displacement at an energy
lower than the peak of the distribution reflects the fact that the
average pulse lies well below the noise.  The figure demonstrates that
there is no evidence for single pulses having energies larger than
$\sim$4.4 times the mean pulse energy.  By contrast, if \psr\ had the
same pulse energy distribution as \ntts, in each of our 12~min
observations, we would have expected over 200 pulses having energy
greater than 5 times the mean.  Thus, PSR J0437$-$4715 does not show
any evidence for ``giant pulses.''

\begin{figure}
\plotone{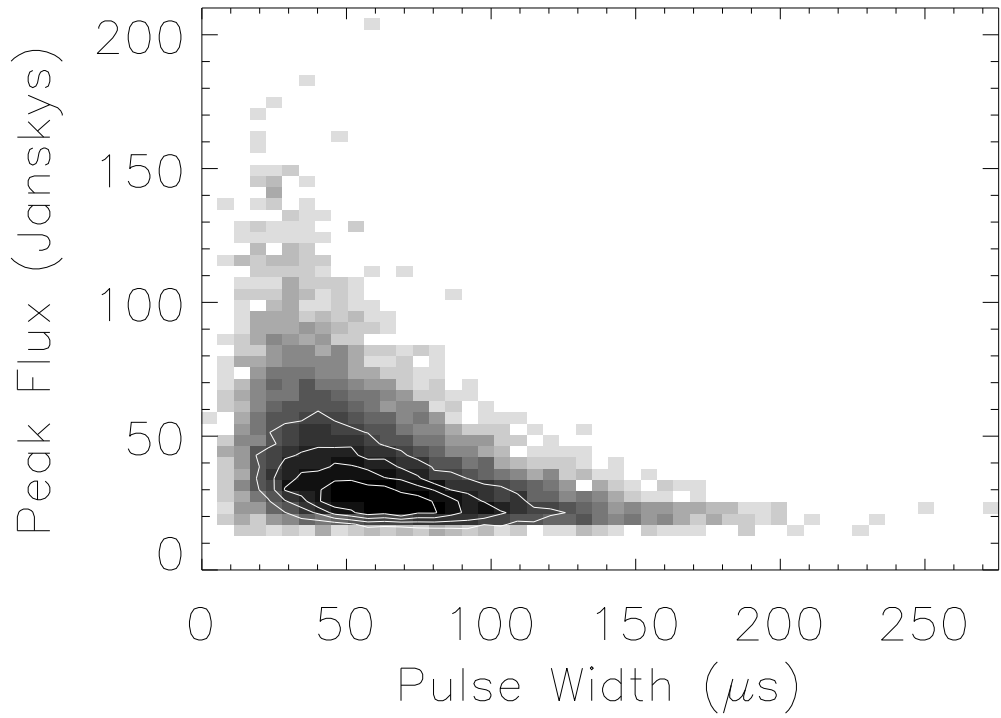}
\figcaption[peakwidthhist.eps]{Pulse peak flux versus pulse width.
Only statistically significant pulses were included in this plot,
hence the lower cutoff near the receiver noise temperature.  Contours
in the plot are 50, 100, 150 and 205 pulses.\label{fig:peakwidthhist}}
\end{figure}

The narrow range of pulse energies exhibited in
Figure~\ref{fig:energyhist}, for a pulsar having very different single
sub-pulse morphologies, suggests there should be a correlation between
pulse peak and width. This correlation is shown in
Figure~\ref{fig:peakwidthhist}.  In the plot, only statistically
significant pulses are included, which explains the lower cutoff in
peak flux at the level of the receiver noise.

\begin{figure}
\plotone{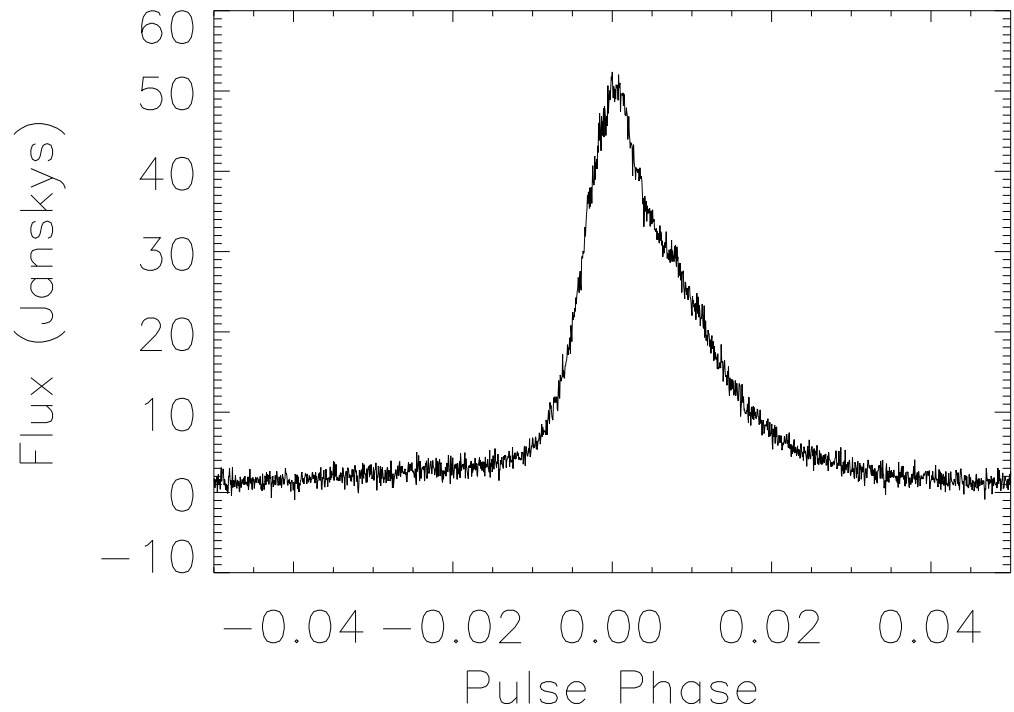}
\figcaption[topfold.eps]{Average profile obtained by folding the 500
single pulses having the highest peak amplitudes in a $\sim$90~s data
span.  The width at half maximum here is 75~$\mu$s.\label{fig:topfold}}
\end{figure}

From Figure~\ref{fig:peakwidthhist}, it is clear that by selecting and
folding only large amplitude pulses, one can
obtain a ``discriminated'' average profile that is considerably
narrower than the standard average profile.  Figure~\ref{fig:topfold}
shows the average profile obtained by selecting the 500 single pulses
having the largest peak amplitudes in a $\sim$90~s span.  Successively
wider average profiles are obtained by lowering that threshold.  The
full width at half maximum of the discriminated average profile shown
in Figure~\ref{fig:topfold} is only $\simeq 75$~$\mu$s,
which should be compared to the full width at half maximum of the
standard average profile, $\simeq 145$~$\mu$s.  This is
interesting, since timing precision improves as the
average profile width decreases; thus discriminated folding could in principle
improve pulsar timing precision.  However, at least in the case of \psr, the
steep spectrum (Fig.~\ref{fig:peakhist}, Eq.~\ref{eq:peakhist}) of
pulse amplitudes precludes timing resolution improvement.

\subsection{Search for Microstructure and Other Preferred Time Scales}

At least eight bright pulsars display structure in their
emission on a time scale shorter than that of the widths of the
average pulse profile and individual subpulses (\cite{mt77}).  Such
``microstructure'' has not previously been detectable in millisecond
pulsars both because of insufficient time resolution of observations,
and because of DM-smearing and multi-path scattering.  Our data on
\psr\ present the first opportunity to detect microstructure in a
millisecond pulsar. Along with classical microstructure we can also
look for evidence of preferred time scales in the emission that may be
due to the presence of primary emitters. The noiselike statistics of
pulsar emission along with the high brightness temperatures lead one
to postulate the existence of many coherent emitters, or primary
emitters, adding up incoherently to form the observed emission
(\cite{mel96}). 

\begin{figure}
\plotone{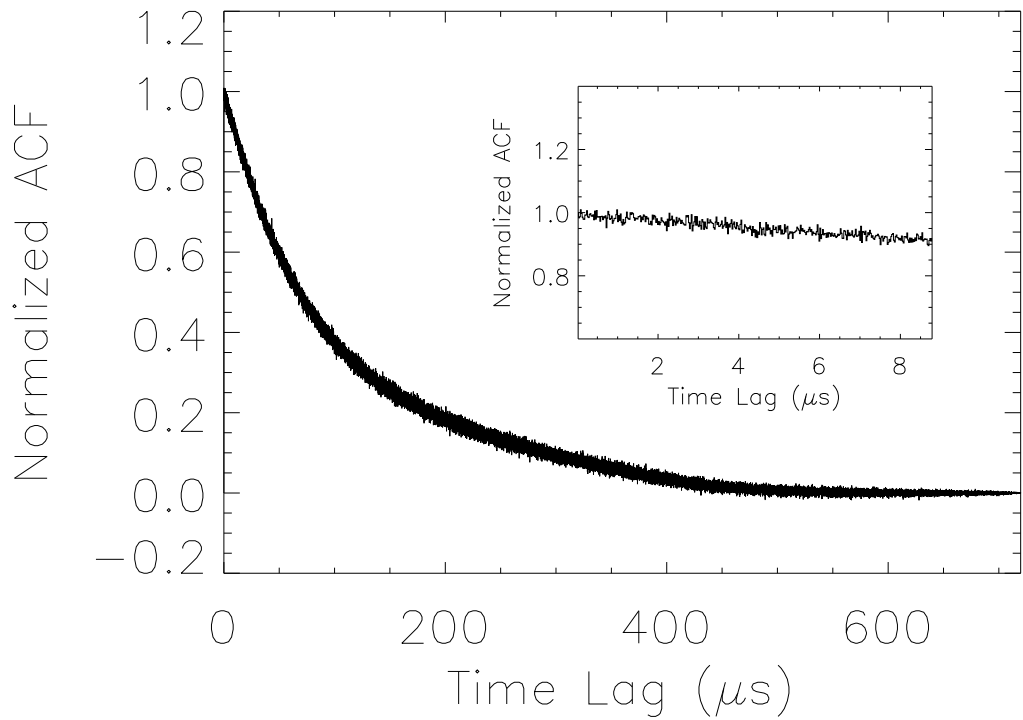}
\figcaption[acf.eps]{Autocorrelation function for $\sim$14,000 consecutive
coherently dedispersed pulses.  The inset is the same data plotted on
a smaller scale. Since no sharp changes in the slope or local maxima
or minima can be seen, there is no evidence for any preferred time
scales other then the subpulse width.\label{fig:acf}}
\end{figure}

We searched for preferred time scales by performing careful
autocorrelation function (ACF) analyses of the incoherently and
coherently dedispersed data. Preferred time scales would present
themselves as rapid changes in the slope of the ACF (i.e. a ``break'')
and/or local minima and maxima in the ACF. In the incoherently
dedispersed data, the time resolution is limited to $\sim$3~$\mu$s; we
find no evidence for microstructure in the $\sim$120,000 incoherently
dedispersed pulses. In order to search with higher time resolution,
ACFs were produced for $\sim$14,000 consecutive coherently-dedispersed
pulses, as well as $\sim$1000 of the largest amplitude coherently
dedispersed pulses. Note that the effective time resolution of the
coherently dedispersed data is limited to $\sim$200~ns by DM
uncertainties and non-ideal filter response. No structure was seen in
the ACF (See Figure \ref{fig:acf}). It is possible that the
microstructures could be individually narrow band. Hence, larger
bandwidths may wash out any preferred time scales. We coherently
dedispersed several sub-bands ranging in width from 6.25 MHz to 781
KHz and centered at 1380 MHz. Again, we found no evidence for
preferred time scales. Note that the highest time resolution achieved
was 80 ns(=$1/2 \times 6.25 \times 10^{6}$).

\subsection{Coherent Radiation Patterns?}

Ables \etal\ (1997) \nocite{amdv97} have reported on single pulse 
observations of \psr\ at
an observing frequency of 326.5~MHz.  They found regularly spaced
peaks in a smoothed arrival time distribution of the largest 500
single pulses in each of their 70~s samples.  These peaks were
spaced by $\sim$20~$\mu$s, much less than their sample time,
102.4~$\mu$s.  They argued, using simulations, that in spite of their
relatively slow sampling, single pulse arrival time uncertainties of
$\sim$5~$\mu$s could be obtained, and hence the fringes could be
resolved.  The fringes implied that certain pulse phases were
preferred for single pulses, which they interpreted as evidence
for a coherent diffractive radiation pattern in the pulsar's emission.
They emphasized that their results were for the largest amplitude
pulses only, which they assumed to be of different origin than
the emission resulting in the standard profile.
Their results predict that with higher time
resolution observations, such as those discussed in this paper, the
average profile obtained by folding the largest amplitude pulses
should show the same fringe pattern. 

Figure~\ref{fig:topfold}, in which we have folded only the highest
amplitude pulses, shows no evidence for any structure, apart from an
asymmetric shoulder on the leading edge, which was seen by Ables \etal\ (1997). We have repeated this analysis by varying the
integration time and pulse threshold, and find similar results.  We
note that our fractional bandwidth is the same as that of Ables \etal\ (1997),
hence the effects of the finite bandwidth should be
identical.  If their apparent fringes were due to a coherent
diffractive radiation pattern, the fringe spacing should scale
with wavelength; thus we would have expected fringe spacings of
$\sim$4.7~$\mu$s, much larger than the 320~ns time resolution of
Figure~\ref{fig:topfold}.  Given the relative sensitivities of our
observations and theirs together with the pulsar spectral index, we
should have been able to observe their ``spikes'' with great ease. 
Thus, we find no evidence to suggest coherent radiation patterns exist. 

It is possible to reconcile the disagreement by postulating that the
``spikes'' have a different radio spectrum than does the emission
forming the average profile.  Ultimately, high time resolution
observations at 326.5~MHz will settle this issue.

\section{Conclusions} \label{sec:concl}

We have presented the first detailed single pulse study of a
millisecond pulsar in which sufficient time resolution was available
to resolve single pulses as they would be seen in the pulsar vicinity.
The similarity of the single pulse properties to those of normal slow
pulsars is remarkable given the dramatically reduced magnetospheric
volume and magnetic field strength of \psr.  Indeed, without being
told the absolute sample rate, it is unlikely that one could
distinguish between this being a millisecond or slow pulsar.  To
summarize, our observations of \psr\ have: resolved individual pulses,
and shown that they have a wide variety of morphologies, with multiple
sub-pulse components not uncommon; shown that individual pulses are in
general broadband; shown that individual pulses can have high linear
polarization; provided no evidence for giant pulses as observed in the
Crab pulsar and PSR B1937+21, nor for pulse nulling or drifting
subpulse phenomena; found structure in the intensity fluctuation
spectrum; revealed a correlation of pulse peak with pulse width so
that the average profile formed from only the highest amplitude pulses
is much narrower than the conventional average profile; not shown any
evidence for micro-structure or preferred time scales $\geq$80~ns;
shown that the emission is consistent with an amplitude modulated
noise model; provided no evidence to support the claim made by Ables
\etal\ (1997) of the detection of coherent radiation patterns.

Because there exists no self-consistent radio emission mechanism
context (see e.g. \cite{mel96}) in which to discuss these results, it
is difficult to say how such models are constrained.  Indeed previous
slow pulsar single pulse studies suffered from the same difficulty.
However, it is often the case that fundamental insights become apparent
when known phenomenon are taken to extremes; this, and improved tape
recording and computer technologies that permit single pulse studies
of millisecond pulsars, provided our motivation in undertaking this
analysis.  Similar studies of other millisecond pulsars may eventually
lead to an understanding of the radio emission mechanism.

\acknowledgements

We thank Jagmit Sandhu, Shri Kulkarni and Don Backer for helpful
conversations.  We thank Matthew Bailes and Jagmit Sandhu for sharing
observing time. This research was supported in part by the National
Science Foundation under grant \#ASC-9318145. VMK received support from
Hubble Fellowship grant number HF-1061.01-94A from the Space Telescope
Science Institute, which is operated by the Association of
Universities for Research in Astronomy, Inc., under NASA contract
NAS5-26555.


\clearpage


\clearpage
\begin{deluxetable}{ccc}
\tablecaption{Epochs of our Parkes 1380 MHz observations of PSR J0437$-$4715.}
\tablehead{\colhead{Observation } & \colhead{MJD at start}  &
\colhead{Duration in minutes}} 
\startdata
1           & 49922.81856   &  12.0    \nl
2           & 49922.82789   &  13.0    \nl
3           & 49923.84874   &  12.4    \nl
4           & 49923.85768   &  13.3    \nl
\enddata
\label{ta:obs}
\end{deluxetable}

\clearpage
\begin{deluxetable}{ll}
\tablecaption{Astrometric and Spin Parameters for \psr\ (\protect\cite{sbm+97}).}
\tablehead{\colhead{Parameter                } & \colhead{Value}}
\startdata
R. A. (J2000) \dotfill & 04$^{\rm h}$ 37$^{\rm m}$ 15\fs748182(4) \nl
Dec.  (J2000) \dotfill & $-47\arcdeg$ 15\arcmin~08\farcs23145(5) \nl
$\mu_\alpha\;{\rm cos}\;\delta$ (mas y$^{-1}$)\dotfill  & 121.34(6) \nl
$\mu_\delta$ (mas y$^{-1}$)\dotfill  & $-$72.50(3)  \nl
Annual parallax (mas) \dotfill & 5.6(8) \nl
Period, $P$  (ms)      \dotfill & 5.75745182525633(6) \nl
Period derivative $\dot P$ ($10^{-20}$) \dotfill &
5.7295(9) \nl
Epoch of period and position (MJD)  \dotfill & 50019.00 \nl
Dispersion Measure (cm$^{-3}$ pc) \dotfill & 2.6469(1)\nl
Binary period, $P_b$ (d) \dotfill & 5.741042353935(350)\nl
$x=a_p {\rm sin}\,i$ (lt-s) \dotfill & 3.36668528(4)\nl
Eccentricity \dotfill & 0.00001920(2)\nl
Long. of periastron, $\omega$ (\arcdeg) \dotfill & 1.793148(20000)\nl
Epoch of periastron (MJD) \dotfill & 50000.49656856(40000)\nl
$\dot x$ ($10^{-12}lt-s~s^{-1}$) \dotfill & 0.082(4)\nl
Timing data span (MJD) \dotfill & 49373 -- 50323 \nl
\enddata
\label{ta:parms}
\end{deluxetable}

\end{document}